\documentstyle[aps,prl,epsf]{revtex}
\draft
\begin{document}
\twocolumn[\hsize\textwidth\columnwidth\hsize\csname@twocolumnfalse\endcsname
\title{Spiral Motion in a Noisy Complex Ginzburg-Landau Equation}
\author{Igor S. Aranson$^{(a)}$, 
Hugues Chat\'e$^{(b)}$, and Lei-Han Tang$^{(c)}$}
\address{
$^{(a)}$Materials Science Division, Argonne National Laboratory, Argonne, 
IL 60439, USA\\
$^{(b)}$CEA --- Service de Physique de l'Etat Condens\'e, 
Centre d'Etudes de Saclay, 91191 Gif-sur-Yvette, France\\
$^{(c)}$Department of Physics and Center for Nonlinear and Complex
Systems, Hong Kong Baptist University, Hong Kong}
\date{\today}
\maketitle
\begin{abstract}
The response of spiral waves to external perturbations
in a stable regime of the two-dimensional complex Ginzburg-Landau 
equation (CGLE) is investigated. It is shown that the spiral core
has a finite mobility and performs Brownian motion when driven by
white noise. Combined with simulation results, this suggests that
defect-free and quasi-frozen states in the noiseless CGLE are unstable 
against free vortex excitation at any non-zero noise strength. 
\end{abstract}
%%%%%%%%%%%%%%%%%%%%%%%%%%%%%%%%%%%%%%%%%%%%%%
\vskip1pc]
\narrowtext
Vortices play a central role in the
ordering of two-dimensional equilibrium systems characterized
by a complex scalar order parameter, e.g., planar magnets
and Bose-condensates \cite{XY}. They also arise in various 
nonequilibrium situations, e.g.,
the Belouzov-Zhabotinsky reaction, certain regimes of fluid 
flows, and the contraction of heart muscles \cite{YK,ATW,MCCPCH,BZ}.
Their long-time motion is determined by their coupling
to the slow modes of the system, and hence is expected
to acquire a universal character.

A distinctive but common feature exhibited by a vortex  in 
a nonequilibrium oscillating state is the emission of spiral waves 
which change the oscillating frequency of the entire system.
This behavior is seen, e.g., from solutions of the generic model 
for these situations, the complex Ginzburg-Landau equation (CGLE) 
\cite{YK,MCCPCH}:
\begin{equation} 
\partial_t a = a - (1+ic ) |a|^2 a + (1+ib)\nabla^2 a,
\label{cgle}
\end{equation}
where $a({\bf x},t)$ is a complex scalar field, and $b$ and $c$ are 
real numbers. Spiral waves are observed for any $b\neq c$.
In certain regimes of the parameter space $(b,c)$,
spiral defects are spontaneously generated and undergo violent
motion, while in other cases they are quite static
and lock into a quasi-frozen structure \cite{HCPM}.
 
In this Letter, we derive mobility relations for a single spiral
in the regime of Eq. (\ref{cgle}), where the uniformly oscillating
state as well as spiral wave excitations are linearly stable. 
We show that, in contrast to the equilibrium XY model (which
corresponds to $b=c=0$), a vortex defect possesses a nonzero
mobility and responds only to perturbations which fall within
a distance $\xi$, the ``screening length'', from the spiral core.
When a noise source is present, the above property yields a
diffusion constant $D$ for the spiral core proportional to the noise
strength. The diffusion constant $D$ is calculated through a
numerical scheme with the result well in accord with direct simulations
of a noisy CGLE.
Secondly, we present simulation results to show that 
both the defect-free state and the frozen-defect state of 
the noiseless CGLE disappear when the (thermal) noise is introduced.
These findings are rationalized in the context of noise-driven vortex
diffusion and thermally-activated nucleation of vortex-antivortex pairs.
We believe the analysis will help toward understanding various other regimes
exhibited by the noiseless CGLE, in particular, the melting of the
frozen-defect state into the defect-turbulence state.

A single-spiral solution to Eq.~(\ref{cgle}) is well documented
\cite{PSH,AKW}. For a single-charged vortex centered
at ${\bf x}_0$, the solution takes the form,
\begin{equation} 
a_s({\bf x},t)= F(r) \exp[ i(\theta + \psi(r) - \omega t)],
\label{spir} 
\end{equation}
where $r=|{\bf x}-{\bf x}_0|$ and $\theta$ is the polar angle
measured from the vortex core.
Far away from the core, the solution approaches a  plane wave with
$\psi(r)\simeq kr$, where the asymptotic wavenumber $k$ is related 
to the rotation frequency as $\omega=c+(b-c)k^2$.
The dependence of $k$  on 
$b$ and $c$ is known analytically for $|b-c|  \ll 1$  and $|b-c| \gg 1$,
e.g. $k\simeq - c^{-1} \exp(-\pi/|2c|)$ for $b=0$ and $|c|\ll 1$ \cite{PSH}. 

In the following discussion, we shall limit ourselves to
the case $b=0$ and $c\neq 0$.
This choice is partly motivated by a desire for simplicity, but it also offers 
an interesting framework: all plane waves 
of wavenumber $k<k_c(c)=(3+2c^2)^{-1/2}$ are linearly stable in this case,
including the homogeneous state $a=\exp (-ict)$ at $k=0$.
For $c<c_i\simeq 1.08$, an isolated spiral is stable and has a wavevector
$k<k_c$ \cite{AKW1}.
In the case $b=c=0$, Eq.~(\ref{cgle}), in the presence of noise, 
reduces to the Ginzburg-Landau equation for a superfluid (or a planar magnet),
allowing the crossover between the equilibrium
XY model and the nonequilibrium CGLE to be explored.
In addition, increasing $c$, solutions of (\ref{cgle}) show a transition
from the quasi-frozen state
to the ``defect turbulence'' state \cite{HCPM}, which can then 
be probed in the presence of noise.

We now consider the response of the spiral to 
a weak additive broad-band noise  $\eta({\bf x},t)$,
\begin{equation}
\partial_t a = a - (1+ic ) |a|^2 a + \nabla^2 a +\eta({\bf x},t).
\label{cgle-1}
\end{equation}
To the linear order in $\eta$, we can write the perturbed solution
in the form,
\begin{equation}
a({\bf x},t)=a_s({\bf x},t)
+\int d{\bf x}' dt' g_s({\bf x},t;{\bf x}',t')\eta({\bf x}',t'),
\label{formal-sol}
\end{equation}
where $g_s({\bf x},t;{\bf x}',t')$ is the Green's function
for the CGLE linearized around $a_s({\bf x},t)$.
Although the exact form of $g_s$ can only be obtained
numerically, it is instructive to examine its behavior
when ${\bf x}'$ is far away from the spiral core. 
%how does a distant perturbation influence the core motion?
This can be done as follows.

In the far field, the spiral wave is very close to
a planewave with a wavevector ${\bf k}=k\hat{\bf e}_r$.
Correspondingly, the inhomogenous Green's
function $g_s$ at large distances can be approximated
by the Green's function $g_p$ of the planewave.
The planewave Green's function can be decomposed into
contributions from two types of modes, the amplitude modes
with a time constant of order one, and the soft phase modes.
Only the latter part survives at long times, yielding the
asymptotic expression,
\begin{eqnarray}
g_p({\bf x}+{\bf x}_0,t+t_0;{\bf x}_0,t_0)&&
\sim {\exp[i({\bf k}\cdot{\bf x}-\omega t)]
\over 4\pi t\alpha_\parallel^{1/2}}\nonumber\\
&&\times\exp\Bigl[-{x_\perp^2\over 4t}
-{(x_\parallel-v_g t)^2\over 4\alpha_\parallel t}\Bigr],
\label{green-plane}
\end{eqnarray}
where $x_\parallel$ and $x_\perp$ are parallel and perpendicular
components of ${\bf x}$ with respect to ${\bf k}$, and 
$\alpha_\parallel=1-2(1+c^2)k^2/(1-k^2).$
(Convective  instability sets in at $\alpha_\parallel=0.$)
The group velocity of the plane-wave is given by
${\bf v}_g=\nabla_{\bf k}\omega=-2c{\bf k}$.

Equation (\ref{green-plane}) implies that a localized perturbation
spreads diffusively but, at the same time, its center travels at
the group velocity ${\bf v}_g$. 
Equating $v_gt$ with the diffusion length $t^{1/2}$,
we obtain a  length $\xi\simeq v_g^{-1}\simeq |ck|^{-1}$.
This is the decay length of disturbances 
in the ``upstream'' direction (i.e., against
${\bf v}_g$), beyond which the influence becomes exponentially small
at any time.
Thus only perturbations within the distance $\xi$
from the spiral core can have a significant influence on its motion.
This is in contrast with the equilibrium XY model ($c=0$ and
$v_g=0$), where the effect of disturbances far away decays only
algebraically with the distance.

A quantitative analysis of the spiral motion under the
perturbation $\eta$ can be done by decomposing
the correction term in (\ref{formal-sol}) into two parts:
(i) shape change of the spiral wave, and
(ii) translation ${\bf x}_0\rightarrow
{\bf x}_0'$ and shift of the overall phase
$\psi_0\equiv\psi(0)\rightarrow \psi_0'$. 
Since the unperturbed spiral wave
is stable, the first type of response decays in time
when the perturbation is switched off. On the other hand,
the second type of response corresponds to the zero modes
of the unperturbed spiral solution and hence does not decay away.
Under an infinitesimal translation of the core position
${\bf x}_0=(x_0,y_0)$, the spiral solution changes to
$a^\prime_s=a_s+x_0w_x+y_0w_y$,
where $w_x$ and $w_y$ are the translational zero modes in the
$x$ and $y$ direction, respectively. They are solutions of
Eq. (\ref{cgle}) linearized around a steady spiral, Eq. (\ref{spir}).
In the presence of the noise  $\eta$, 
the equations for the translation zero modes assume the form: 
\begin{equation}
\dot{x}_0=\zeta_x,\quad 
\dot{y}_0=\zeta_y,
\label{diff-eqn}
\end{equation}
where $\zeta_x$ and $\zeta_y$ are projections of $\eta$ 
on the zero modes $w_x$ and $w_y$, respectively.

While the form of the functions $w_x$ and $w_y$ can be
obtained directly by differentiating (\ref{spir}), the calculation
of $\zeta_x$ and $\zeta_y$ is quite difficult.
This is because the operator ${\cal L}$ which defines the linearized
equation around a spiral solution is non-hermitian except
at $c=0$, so the eigenfunctions $u_\alpha, \alpha=x,y$,
of the adjoint operator ${\cal L}^\dagger$
are in general different from the eigenfunctions $w_\alpha$.
For $c \sim 1 $, we have developed a numerical scheme \cite{alg} to determine
$u_\alpha$, details of which will be reported elsewhere \cite{act}.
Quite generally, $u_\alpha(r,\theta)$ decays exponentially
at large distances from the spiral core \cite{AKW}. From the projection
formula,
\begin{equation}
\zeta_\alpha={\int d^2x u_\alpha^\dagger\eta\over
\int d^2 x u_\alpha^\dagger w_\alpha},
\label{zeta-def}
\end{equation}
we see that only the part of the perturbation $\eta$ sufficiently
close to the core contributes to $\zeta_\alpha$, in accord with
the qualitative analysis developed above.
The exponential decay of $u_\alpha$ also ensures the convergence of 
the integral on the denominator, yielding a finite ``mobility'' 
to the spiral core. This contrasts with the $c=0$ case,
where the integral diverges, and the vortex mobility decreases
logarithmically with the relevant length scale.

We now apply the above discussion to the special case where
the external perturbation takes the form of a weak, uncorrelated
white noise with zero mean and correlators,
\begin{equation} 
\langle\eta_\beta({\bf x}, t) \eta_\gamma({\bf x}', t^\prime)\rangle
=2 T\delta_{\beta\gamma}\delta(t-t^\prime) \delta({\bf x}-{\bf x}^\prime),
\label{noise-correlator}
\end{equation}
where $\beta, \gamma$ specify real and imaginary parts of $\eta$ and
$T$ characterises the noise strength.
Solving Eq. (\ref{diff-eqn}), we obtain a diffusion law for the
vortex core at long times, 
\begin{equation}
\langle [{\bf x}_0(t)-{\bf x}_0(0)]^2\rangle=4Dt.
\label{diff-cnst}
\end{equation}
For small $T$, the diffusion constant is $D=\mu T$, with the mobility
given by
\begin{equation}
\mu={\int d^2 x u_x^\dagger u_x
\over |{\int d^2 x u_x^\dagger w_x}|^2}.
\label{mobility}
\end{equation}

We have evaluated Eq. (\ref{mobility}) for several different
values of $c$ using the numerically determined eigenfunctions
$u_\alpha$ and $w_\alpha$, and have also performed 
Langevin simulations of the noisy
CGLE (\ref{cgle-1}). In the simulations, the core coordinates
of an initially prepared spiral
was followed in time at various noise strengths $T$.
Brownian motion of the core was observed. In Fig. 1(a), we
show the simulation data for the mean-square deviation of the
core coordinates versus time for $c=1.4$ and various values
of temperature $T$. From the slope of the curves, we determine the
diffusion constant $D$, which is found to vary linearly with $T$
at small $T$. In Fig. 1(b), we plot the mobility $\mu$ against
the parameter $c$, where filled squares correspond to data
obtained from simulations, and open circles data from
direct calculations using Eq. (\ref{mobility}). 
The agreement
between the two sets is apparent.

\begin{figure}
\narrowtext
\vspace{0.3cm}

\centerline{\epsfxsize=5.3truecm
\epsffile{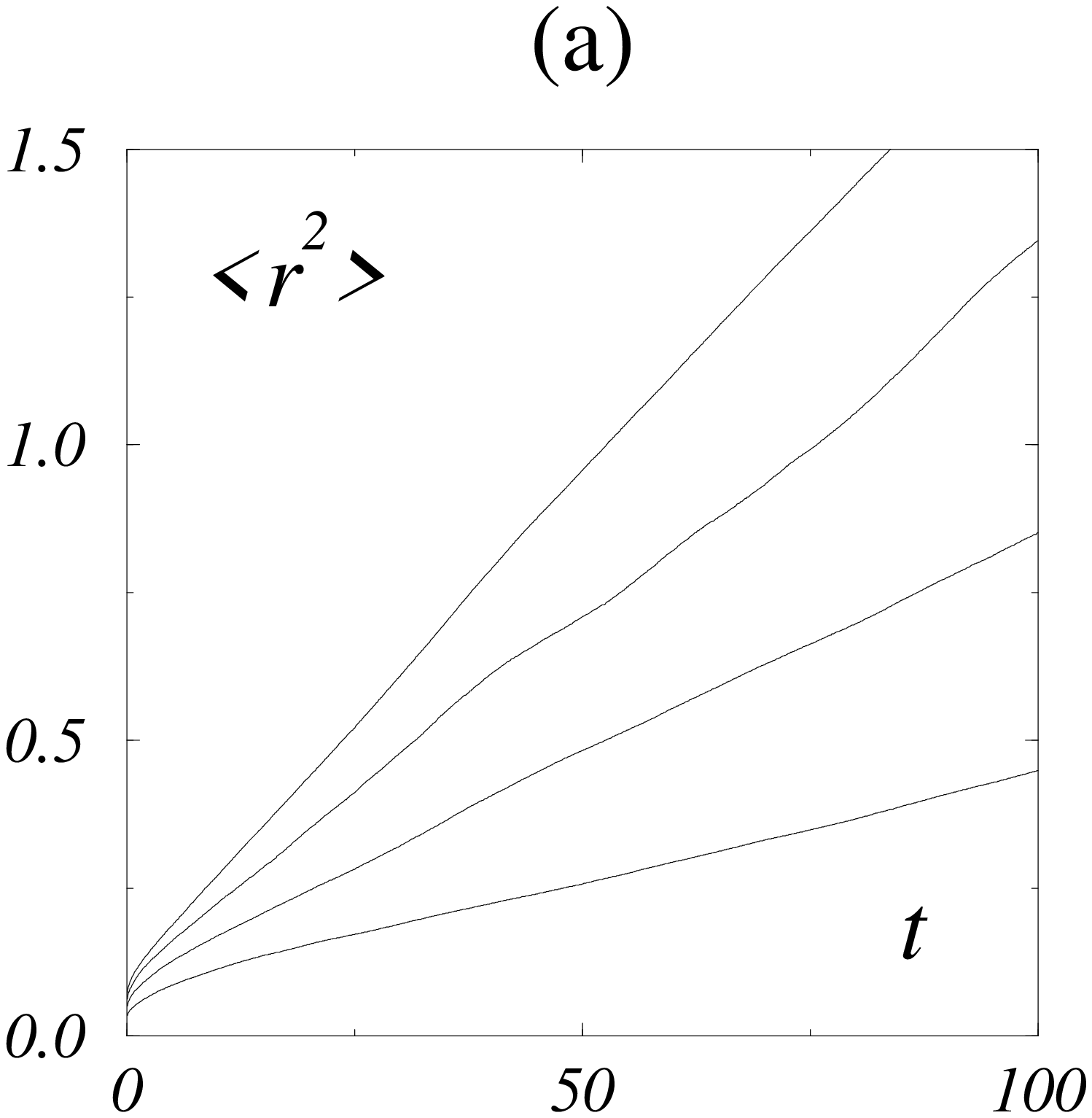}
\hspace{-1.2cm}
\epsfxsize=5.3truecm
\epsffile{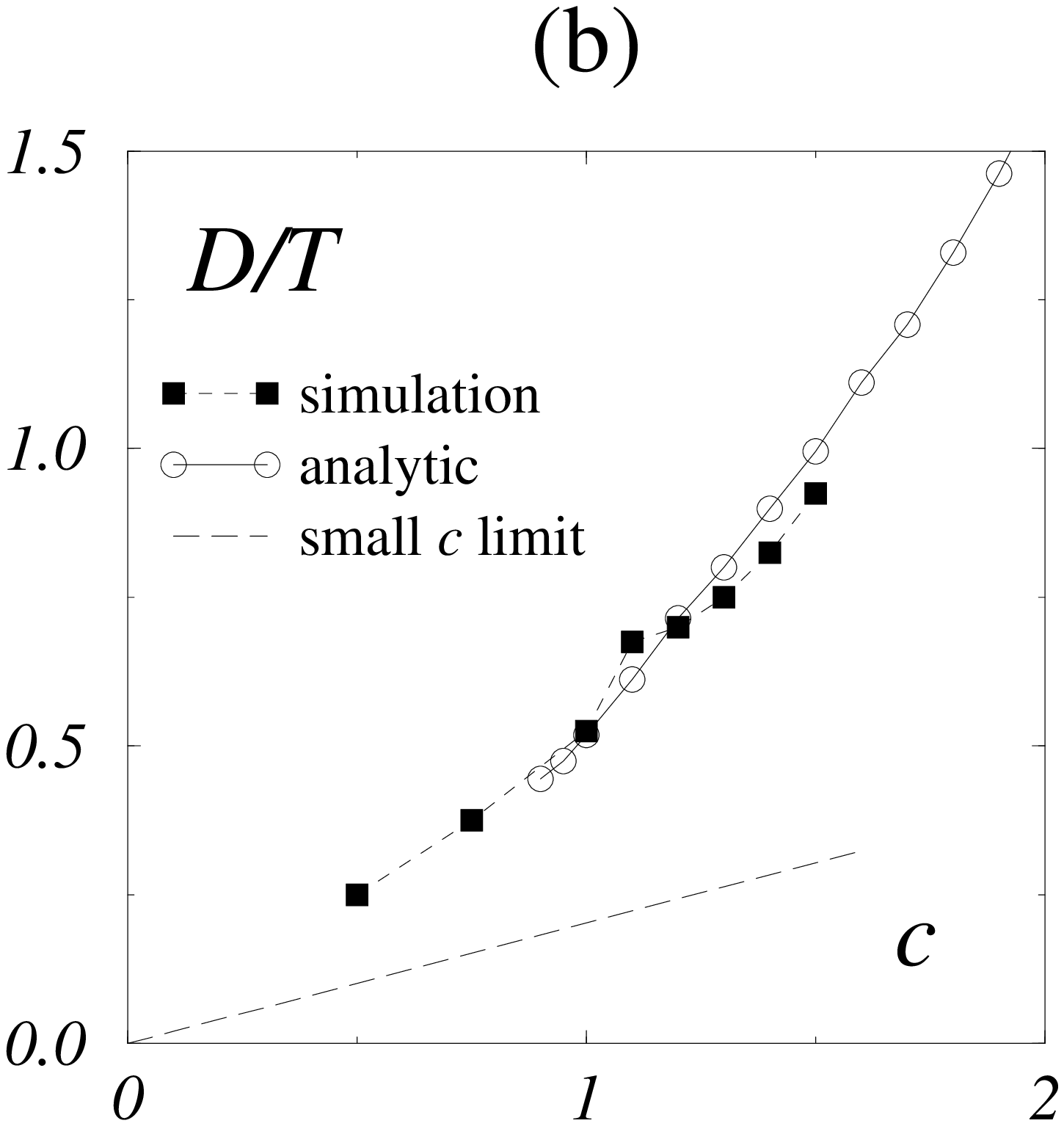}}
\vspace{-0.5cm}
\caption{Noise-driven spiral diffusion: (a) mean-square deplacement
$\langle r^2 \rangle = \langle [{\bf x}_0(t)-{\bf x}_0(0)]^2\rangle$
vs. time for $c=1.4$, $T=0.001$, 0.002, 0.003, and 0.004 (from bottom to top).
Data obtained from the trajectory, over $\Delta t \sim 10^4$,
of a spiral initially centered in 
a disk of radius 128 with no-flux boundary conditions.
(b) mobility $\mu=D/T$ vs. $c$.}
\label{fig1}
\end{figure}

In the special case $c=0$, the problem reduces to that of
vortex diffusion in an equilibrium XY model, which has been
studied extensively in the past. In this case,
the linear operator ${\cal L}$ becomes hermitian, and
$u_\alpha=w_\alpha$. 
For $|c|\ll 1$, we have analyzed the
eigenvalue equation for $u_\alpha$ in detail,
and found that
$u_\alpha\simeq w_\alpha$ at distances up to $r\sim \xi$ before
it switches over to exponential decay at $r>\xi$. Using this property
and the exponential divergence of $\xi$ for  $c \to 0$, we obtain
\begin{equation}
\mu\simeq 2 c / \pi^2 \;.
\label{small-c}
\end{equation}
We have checked that this small $c$ behavior is consistent
with the trend exhibited by the numerical data shown in Fig.~\ref{fig1},
though the next order term is significant when $c$ reaches
a value of order one.

The discussion presented above on the diffusivity of a single
vortex serves as a useful basis for understanding the behavior
of the noisy CGLE (\ref{cgle-1}) under more general initial
conditions. Specifically, we have examined numerically
the behavior of the system under two representative initial
conditions: (a) a uniform, defect-free state with $a=1$,
and (b) a random initial state with $|a|\ll 1$.
(In the latter case the system initially develops into
a configuration with a high density of vortices.)
It has been shown that, in the absence of noise, 
case (a) is stable and the system remains synchronized
(i.e., with bound or no phase fluctuations) at all times,
while case (b) evolves into a ``quasifrozen-defect'' state
for $c<c^\ast\simeq 1.68$ and a ``defect-turbulence'' state 
for $c>c^\ast$ \cite{HCPM}.
Our simulations show that, when noise is present, 
the distinction disappears and the system evolves into
a single steady-state with a finite density of vortices.

\begin{figure}
\narrowtext
\centerline{
\epsfxsize=7.5truecm
\epsffile{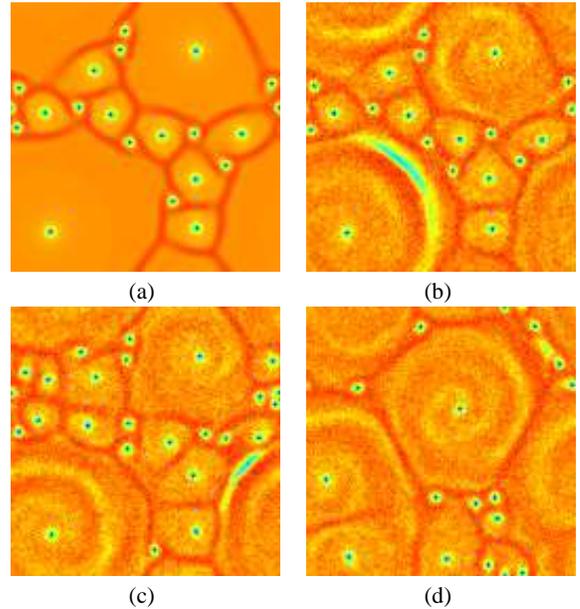}
}
\caption{Snapshots of $|a|$ during the breakdown of a blocked configuration 
at $c=1.4$. System of linear size $L=128$ with periodic boundary conditions.
(a): blocked state ($T=0$).
(b,c): initial stages ($T=0.002$, $t\sim 500, 700$); note the $|A|$ modulations
emitted by the cores of the large spirals.
(d): in the asymptotic state ($T=0.002$, $t \sim 5000$), spirals diffuse; 
their maximal size is fixed by the strength of 
the convective instability and the noise level.}
\label{fig2}
\end{figure}

Figure 2 shows the breakdown, under weak noise ($T=0.002$),
of an initially-blocked configuration [Fig.~\ref{fig2}(a)] 
obtained at $c=1.4$ by simulating the noiseless equation with a random 
initial condition. Clear spots with a black center indicate the depression of
$|a|$ at the vortex cores.
Switching on the noise, we see the diffusive motion of
individual vortices, creation of new vortices ---in this case, through
noise amplification by a convective instability
[clear belt in the lower left corner in Fig. 2(b)]---, and 
annihilation of vortex-antivortex pairs. For weaker noise, the whole process
occurs on larger timescales, reaching an asymptotic state with larger
spirals.

\begin{figure}
\narrowtext
\vspace{0.3cm}
\centerline{\epsfxsize=5.3truecm
\epsffile{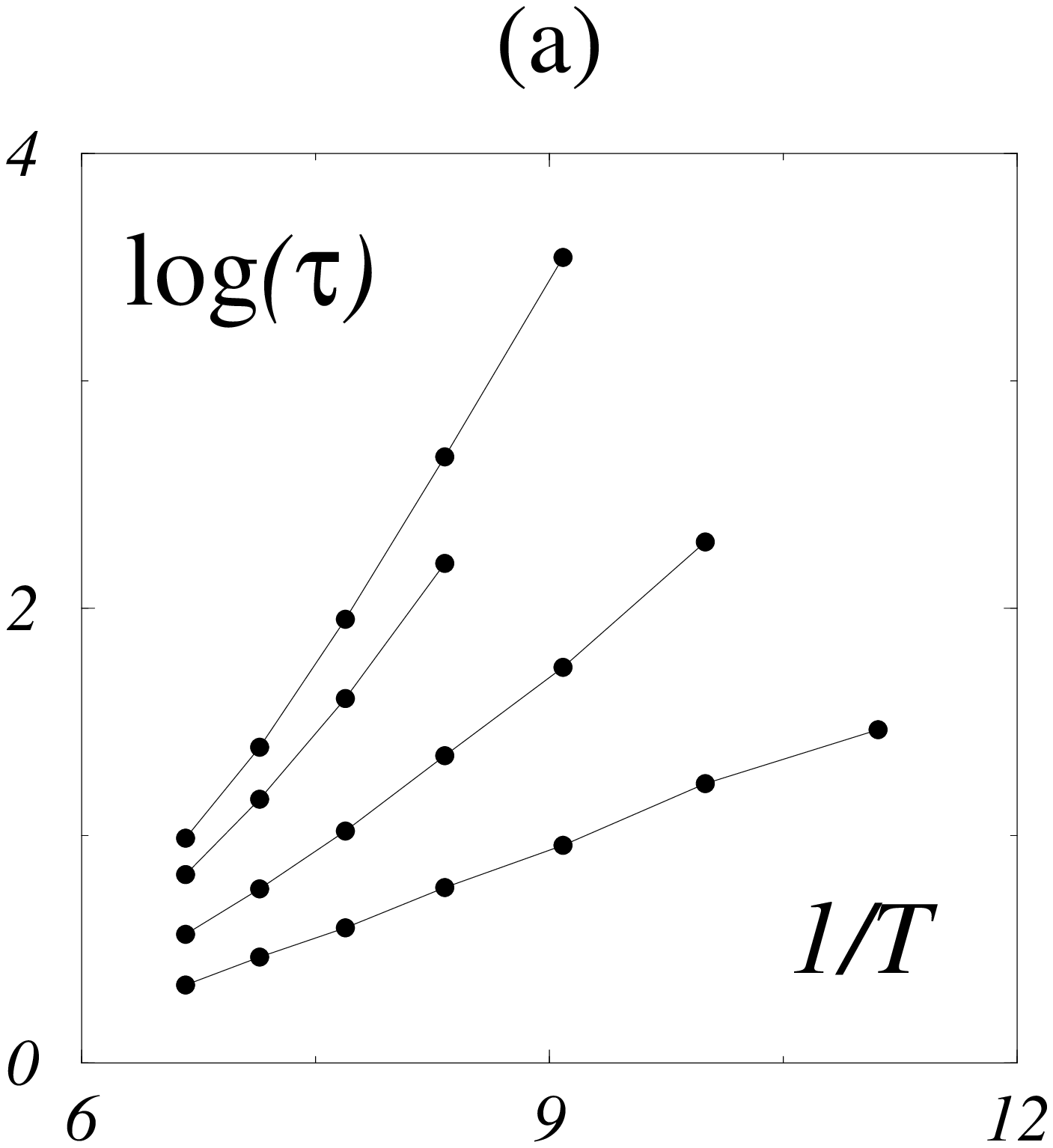}
\hspace{-1.2cm}
\epsfxsize=5.3truecm
\epsffile{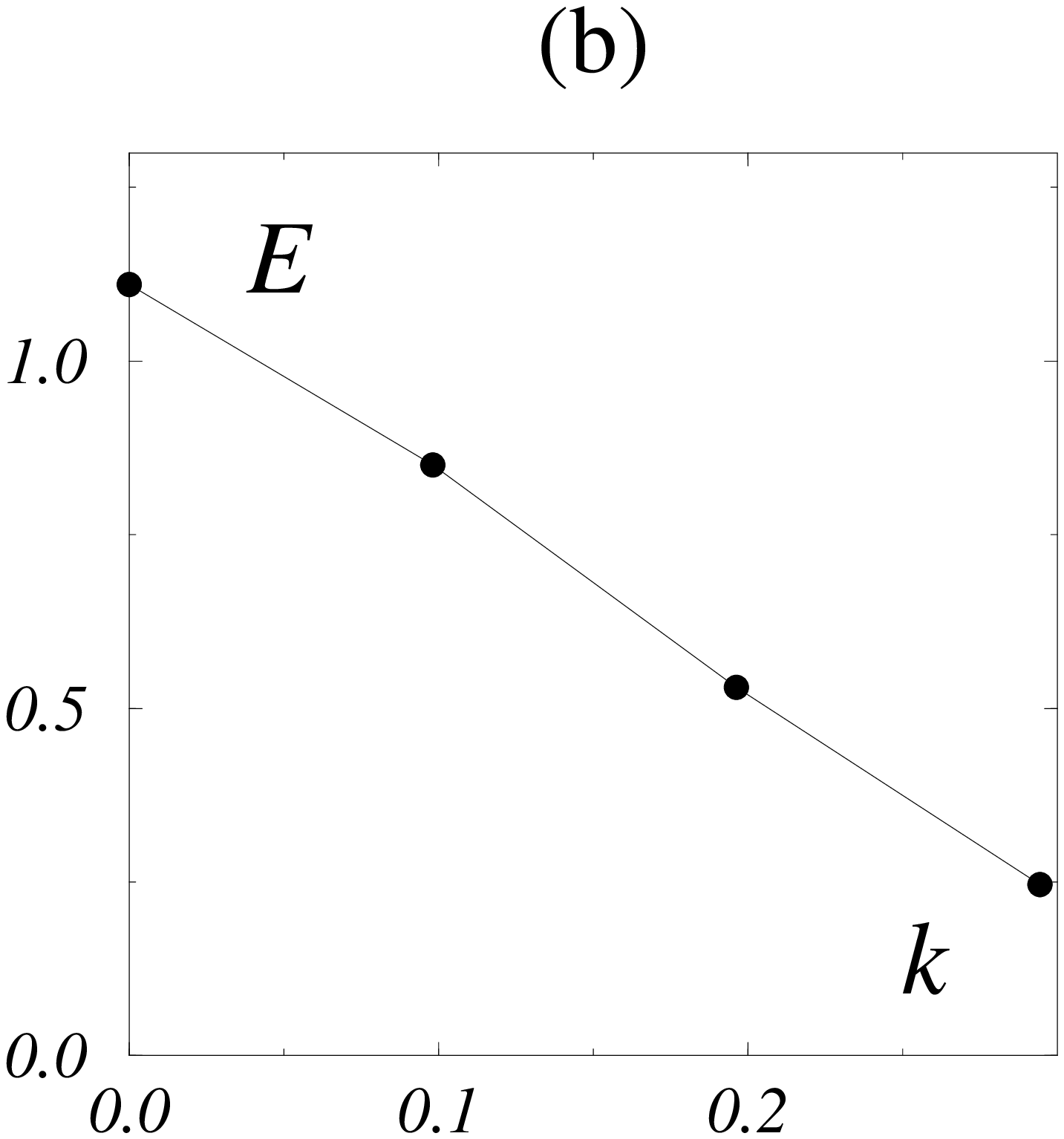}}
\vspace{-0.5cm}
\caption{(a) Logarithm of the average nucleation time $\tau$
of a thermal pair of vortices vs. $1/T$ for flat, tilted
initial conditions with mean phase gradient $k$. 
From top to bottom: increasing $k$, from $k=0$ to $k\simeq 0.3$.
Each point corresponds to 500 runs of a system of size
$L=128$ at $c=1.25$. (Even though $c>c_i$, the convective instability
has no influence on these results.)
(b) Nucleation energy $E$, as measured from the small $T$ behavior of (a)
($\tau \propto \exp (-E/T)$), vs. $k$.}
\label{fig3}
\end{figure}

For $c<c_i$, the convective instability is absent and 
vortex-antivortex pairs are created  solely through
a ``thermal'' activation process. (Although this mechanism
is also present for $c>c_i$, it does not play a dominant role
when the noise is weak.)
Since the amplitude of $|a|$ has to be zero at the vortex core,
nucleation of a pair requires a rare fluctuation in the noise
amplitude, unless it happens next to the core of an existing
vortex. Indeed, most of the newly created ``thermal''  pairs 
are found next to existing vortices in our simulations.
When the noise is weak, the probability for the creation
of a thermal pair can be argued to be proportional to an exponential factor,
$\exp(-E/T)$, where the ``activation energy'' $E$ is expected
to be lower when next to an existing vortex than in other regions of
the system. This is borne out by our simulation data,
which show the exponential dependence on $T$ of the 
ensemble-averaged  time $\tau$ necessary for the nucleation of a 
thermal pair from defect-free, tilted
initial conditions  (Fig.~\ref{fig3}a). We also find an almost linear
variation of $E$ with phase gradient  (Fig.~\ref{fig3}b). 
Moreover, $E$ is found not to be sensitive to the value of $c$ \cite{act}.

In the case of the equilibrium XY model, 
below the Kosterlitz-Thouless transition temperature\cite{XY}, 
vortices of opposite charge attract one another through a 
logarithmic Coulomb potential and remain bounded in pairs. 
For $c\neq 0$, however, this attraction is cut off
at the distance $\xi$, beyond which the interaction becomes
exponentially weak \cite{AKW}. 
Thus, vortices become ``free'' when their separation
is larger than $\xi$. 

%Below the critical value $c_i$ for the convective instability,
For $c\ne 0$ the diffusive motion of individual vortices makes it possible
for the nucleated vortices to break away from one another. 
(Most of them are short-lived, but a fraction do become ``free'' \cite{act}.)
On the other hand, it also enables two distant,
oppositely-charged vortices to move into the interaction range $\xi$ 
and annihilate each other.
The steady-state density $n$ of free vortices is thus controlled
by both the ``pumping'' rate $p$
of creating vortex pairs of size $\xi$,
and the diffusion constant $D$
which limits the rate of annihilation. 
On a mean-field level, we can write: 
$p=Dn^2$.
When $n$ is small, we may write $p(n)=p_0+p_1n$, where $p_0$
is the nucleation rate of pairs in the far field, while
$p_1$ is the nucleation rate next to a free vortex.
The resulting equation has a unique positive solution,
corresponding to a unique steady-state. 
Our numerical experiments agree with this conclusion: 
from both synchronized and random initial conditions, a single steady-state
seems to be reached \cite{act}.

For $c>c_i$,
the asymptotic state possesses a nonuniform density of vortices.
There are two relevant length scales in this case:
the ``noise''  length $ \sim \log (1/T)$  given by noise 
amplification condition,  
which determines 
the size of the large spirals, and the length scale $\xi  $
below which vortices of opposite charge annihilate each other. 
(At such large values of $c$,
$\xi$ becomes comparable to the size of the vortex core.)
The system is better described as a gas of large,
well-developed spirals with a texture of precipitated vortices
at their boundaries [Fig.~\ref{fig2}(d)]. 

In conclusion, we have shown that vortices in the CGLE
possess a finite mobility and diffuse under the influence of
an external white noise. The diffusion constant at small
noise strength is calculated through a numerical scheme which 
compares favorably with direct simulations of a noisy CGLE.
Our results indicate that the previously observed frozen-defect
state of the noiseless CGLE is unstable against noise, and
hence it does not possess the kind of metastability exhibited by
a typical glassy state. Noise also leads to the creation of
free vortices through either an activated process or a dynamic
instability. The steady-state of the system becomes unique and
contains a finite density of free vortices, in contrast to the
behavior of the equilibrium XY model below the Kosterlitz-Thouless
temperature.

HC and LHT wish to thank, respectively, 
the Hong Kong Baptist University and the Materials Science
Division, Argonne National Laboratory, for hospitality, 
where part of this work was performed.
The work of IA was  supported by the U.S. Department
of Energy under contracts W-31-109-ENG-38 (IA) and
ERW-E420 and the NSF, Office of STC
under contract No. DMR91-20000.


\begin{references}
\bibitem{alg} To determine the eigenfunctions $u_\alpha$ and $w_\alpha$, 
we solved simultaneously, using a matching-shooting algorithm,  
three  nonlinear equations for the 
unperturbed spiral + eight linear equations for the
adjoint translation modes.
\end{references}
\end{document}